\documentclass[showpacs,aps,twocolumn]{revtex4-1}

\usepackage{amsmath}
\usepackage{amssymb}
\usepackage{amsthm}
\usepackage{bbm}
\usepackage{mathrsfs}
\usepackage{bm}
\usepackage{graphicx}

\newcommand{ \rhot }{ \varrho }
\newcommand{ \rhosys }{ \rho }
\newcommand{ \Ht }{ {\cal H} }

\newcommand{ \Jt }{ {\cal J} }

\newcommand{ \sL }{\mathbb{L}}
\newcommand{ \sK }{\mathbb{K}}

\newcommand{ \sM }{\mathbb{M}}
\newcommand{ \sG }{\mathbb{G}}
\newcommand{ \sD }{ \mathbb{D} }

\newcommand{ \sW }{\mathbb{W}}

\newcommand{\rhotot}{\varrho}

\newcommand{ \sLsys }{\mathbb{L}_{\rm sys}}
\newcommand{ \sLtot }{\mathbb{L}_{\rm tot}}

\newcommand{ \bK }{\boldsymbol{K}}
\newcommand{ \mJ }{\boldsymbol{J}}

\newcommand{\bH}{\mathbb{H}}

\newcommand{\Tr}{\mbox{Tr}}                                        
\newcommand{\nn}{\nonumber}

\newcommand{\be}{\begin{equation}}
\newcommand{\ee}{\end{equation}}

\newcommand{\inpr}[2]{(#1|#2)}

\begin{document}

\title{A generic map from non-Lindblad to Lindblad master equations}

\author{Michael R. Hush$^{1,2}$}
\author{Igor Lesanovsky$^1$}
\author{Juan P. Garrahan$^1$}
\affiliation{$^1$School of Physics and Astronomy, University of Nottingham, Nottingham, NG7 2RD, UK}
\affiliation{$^2$School of Engineering and Information Technology, University of New South Wales at ADFA, Canberra, ACT 2600, Australia}
\date{\today}

\begin{abstract}
Many current problems of interest in quantum non-equilibrium are described by time-local master equations (TLMEs) for the density matrix that are not of the Lindblad form, that is, that are not strictly probability conserving and/or Markovian.  Here we describe an generic approach by which the system of interest that obeys the TLME is coupled to an ancilla, such that the dynamics of the combined system-plus-ancilla is Markovian and thus described by a Lindblad equation.  This in turn allows us to recover the properties of the original TLME dynamics from a physical unravelling of this associated Lindblad dynamics.  We discuss applications of this generic mapping in two areas of current interest.  The first one is that of ``thermodynamics of trajectories'', where non-Lindblad master equations encode the large-deviation properties of the dynamics, and we show that the relevant large-deviation functions (i.e.\ dynamical free-energies) can be recovered from appropriate observables of the ancilla.  The second one is that of quantum filters, where we show tracking a quantum system undergoing a continuous homodyne measurement with another quantum system of the same size will inherently be inefficient in our framework.
\end{abstract}

\maketitle

\noindent
\section{Introduction} 
A central result in the theory of open quantum system is due to Lindblad \cite{Lindblad1976,Gorini:1976,Plenio1998,Breuer2002,Gardiner2004} who proved that the general form for the quantum master equation (QME) for the density matrix of a quantum Markovian system is,
\begin{equation}
\dot{\rhot} = -i[\Ht,\rhot] + \sum_{i=1}^{N} \Jt_i\rhot \Jt^\dag_i - \frac{1}{2} \{ \Jt^\dag_i \Jt_i, \rhot \} \equiv \sL(\rhot) ,\label{eqn:introgenMarkov}
\end{equation}
where the self-adjoint Hamiltonian $\Ht$ generates the coherent part of the dynamics, $\Jt_i$ are a set of $N$ (bounded) {\em jump} operators which encode incoherent transitions \cite{Lindblad1976,Plenio1998,Breuer2002,Gardiner2004}, and $\{ \cdot,\cdot \}$ indicates anticommuator. We allow both $\Ht$ and $\Jt_i$ to possibly be time dependent: when they are not, Eq.~(\ref{eqn:introgenMarkov}) is Markovian, and when they are, Eq.~(\ref{eqn:introgenMarkov}) is time-dependent Markovian \cite{Breuer2009,Wolf2008}.  A large number of open quantum systems of experimental relevance have been described using this Markovian approximation \cite{Lindblad1976,Plenio1998,Breuer2002,Gardiner2004}, and Eq.~(\ref{eqn:introgenMarkov}) has become the starting point for analysing the dynamics of a quantum system that interacts with a thermal bath or other environments.  

The form Eq.~(\ref{eqn:introgenMarkov}) guarantees positivity of the density matrix, conservation of probability, and the quantum (i.e.\ no memory) Markov property \cite{Wolf2008,Breuer2002}.  Useful theorems have been proved for it, including on the existence of stochastic unravellings \cite{Belavkin1990,Gardiner2004,Wiseman2010}, and on steady state behaviour \cite{Kraus2008,Schirmer2010}. Furthermore, frameworks now exist to synthesise Markovian dynamics, of almost any kind, in experiment: through network synthesis of linear quantum systems \cite{Wang:2012,Nurdin:2009}, or the engineering of quasi-local dissipation \cite{Kraus2008,Verstraete:2009,Diehl:2008,Weimer:2010,Ticozzi:2014}. Thus if the master equation is Markovian, it is frequently possible to realise it with quantum hardware. 

No such framework exists in general for master equations which are \emph{not} of the Lindblad kind.
This paper aims to remedy this problem for general time local master equations (TLMEs) of the form,
\begin{equation}
\dot \rhosys = \sL_{\rm sys}(\rhosys) + B\rhosys + \rhosys C + \sum_{j=1}^{M} D_{j} \rhosys E_j^\dag \equiv \sK(\rhosys) , \label{eqn:genTLME}
\end{equation}
where $\sL_{\rm sys}$ is a Lindbladian, as in Eq.~(\ref{eqn:introgenMarkov}), and $B,C,D_{j},E_{j}$ are arbitrary (bounded) operators which may or may not be time-dependent \footnote{We denote by $\rhosys$ the density matrix of a system evolving according to a TLME, by uppercase Roman symbols the operators that act on its Hilbert space, by $\rhot$ the density matrix of a system evolving according to a Lindblad master equation, and by uppercase calligraphic the corresponding operators.  We use blackboard font for super-operators throughout. \label{footy}}. 
The super-operator $\sK$ defined by Eq.~(\ref{eqn:genTLME}) does not in general preserve positivity or probability of the density matrix $\rhosys$.
 
TLMEs such as Eq.~(\ref{eqn:genTLME}) appear in three important contexts in quantum non-equilibrium: 

\noindent
{\em (i)} One is explicitly {\em non-Markovian systems} \cite{Laine2012,Shibata1980,Chaturvedi1979,Kubo1963}. Understanding non-Markovian dynamics is of current interest as such evolution has been discovered in systems as varied as photosynthetic complexes \cite{Lee:2007,Rebentrost:2011,Mujica-Martinez:2013}, solid-state systems\cite{Coish:2004,Banyai:1995}, quantum optical setups \cite{Liu:2011}, trapped ions \cite{Maniscalco:2004,Budini:2001,Turchette:2000} and cold atoms \cite{Breuer:1999,Hope:2000}. It is sometimes argued that integro-differential equations \cite{Nakajima1958,Zwanzig1960} must be solved to model non-Markovian processes, but this is not true in general. Any non-Markovian process described by a time non-local integro-differential equation can be transformed into a time local master equation (TLME) by introducing an exact backward propagator of the total system into the integro-differential equation \cite{Breuer2004,Chruifmmode:2010}. For very strong couplings the TLME formalism for non-Markovian dynamics may have singularities but the dynamics remains regular \cite{Chruifmmode:2010}. Furthermore, there are also cases where the TLME can provide more accurate predictions than time non-local formalism \cite{Royer1972,Royer1996,Royer2003}. Non-Markovian evolution can result in negative eigenvalues for the density matrix for short times, these fluctuations cannot be captured by Lindblad master equations as they keep the density matrix positive for all times. 

\noindent
{\em (ii)} The second area is full-counting statistics in quantum optics \cite{Gardiner2004} and in mesoscopics \cite{Levitov1996,*Esposito2009}, and the related {\em thermodynamics of trajectories}, the generalization of Ruelle's thermodynamic formalism \cite{Eckmann1985,*Ruelle2004} to many-body stochastic systems, both classical \cite{Merolle2005,Lecomte2007} and quantum \cite{Garrahan2010,Budini2010}.  Here TLMEs encode the large-deviation properties \cite{Touchette2009} of time-integrated observables, and the associated large-deviation functions play the role of dynamical free-energies for ensembles of trajectories \cite{Merolle2005,Lecomte2007,Garrahan2010}. The TLMEs itself does not necessarily conserve the norm of the density matrix, indeed in the long time limit the smallest eigenvalue of the TLME superoperator corresponds to the value of the large deviation function (which can be non-zero). Furthermore when the ``counting'' field $s$ (see \cite{Garrahan2010,Budini2010} for details) is complex, the TLME can generate density matrixes which are non-positive.  

\noindent
{\em (iii)} The third area is quantum feedback and control \cite{Wiseman2010,Gardiner2004}, in particular {\em quantum filters} \cite{Belavkin1990,Wiseman1993,Strunz1999,Bouten2007,Szigeti2009,Szigeti2010,Cui2012,Dotsenko2009,Geremia2003} whose dynamics is also described by TLMEs \cite{Bouten2007}. Quantum filters give an optimal estimate of the state of a quantum system conditioned on a continuous homodyne measurement record. They can be used in real-time with an experiment, and provide an estimate that can be to control the experiment with feedback \cite{Doherty:1999,Combes:2006,Handel:2005,Steck:2004,James:2008,Szigeti:2013}. This has already been experimentally demonstrated in cavity QED experiments \cite{Sayrin:2011} and atomic spin systems \cite{Geremia:2004}. The TLME used in quantum filtration ensure the density matrix is positive, however some filters do not necessarily conserve the norm of the density matrix \cite{Wiseman2010}. 
  
 \begin{figure*}[tb]
\includegraphics[width=\textwidth]{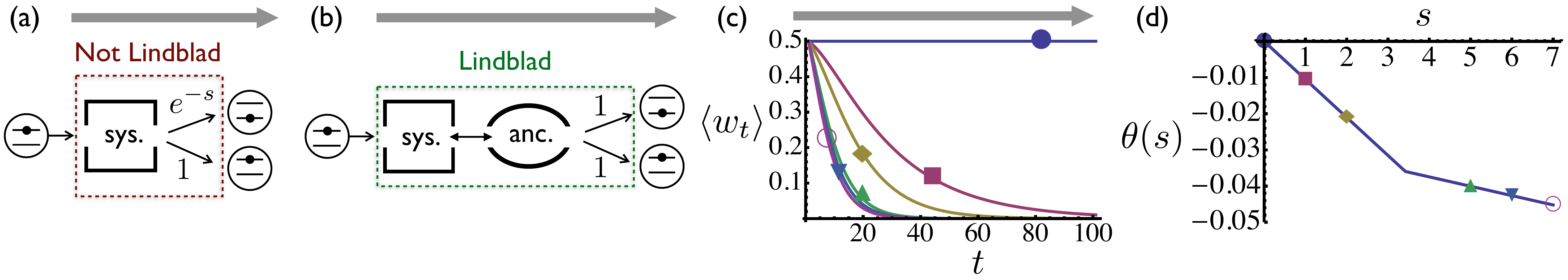}
\caption{(Color online) We consider how to map a system that obeys a TLME Eq.~(\ref{eqn:genTLME}) which is not in Lindblad form, (a), to a hybrid system-ancilla, (b), whose total evolution \emph{is Markovian}, Eq.~(\ref{eqn:introgenMarkov}), which under a \emph{quantum weighted average}, Eq.~(\ref{eqn:quantweight}), reproduces the TLME for the original system. In this case (a) is a micromaser---a cavity pumped by excited two-level atoms---biased by a ``counting'' field $s$. This results in non-Lindblad evolution (see text for details) where the probability to detect a ground state atom leaving the cavity is biased by the factor $e^{-s}$ with respect to unbiased dynamics. Such non-Lindblad evolution is mimicked by a micromaser  coupled to an ancilla. The composite system (b) evolves under a physical Markovian dynamics. We evolve a Markovian quantum jump Monte Carlo of the composite system then extract the dynamics of the embedded non-Lindblad evolution. The decay of the coherence of the ancilla is plotted in (c). The decay rate of the coherence in the Markovian simulation is precisely the large-deviation function $\theta(s)$, which corresponds to the largest eigenvalue of the non-Lindblad TLME. $\theta(s)$ can be extracted in this manner and contains the full counting statistics of the atoms leaving the cavity. In (d) we plot the function of $\theta(s)$ and also the points determined using the decay rates of the coherence in (c). In (d) we can see the micromaser at these conditions $\theta(s)$ has a first-order singularity at $s_c \approx 3.4 \times 10^{-4}$, indicating a dynamical phase transition (for the micromaser it is actually a very sharp crossover) \cite{Garrahan2011,Horssen2012}.}
\label{fig1}
\end{figure*}
 
\section{The Map} 
 
Here we present a generic way to map a system whose dynamics is described by a TLME to one where the same system is coupled to an ancilla, such that the density matrix $\rhot$ of the combined system-plus-ancilla evolves according to a master equation of the Lindblad form.  The evolution of the system under the TLME is recovered via:
\begin{equation}
\rhosys = \Tr_{\rm a}[w_{t}\rhot], \label{eqn:quantweight}
\end{equation}
where $w_{t}$ is an appropriate (possibly time-dependent) operator on the ancillary space and $\Tr_{\rm a}$ indicates trace over the ancillary Hilbert space.  We achieve this using a novel framework called \emph{quantum weighting}, named due to its analogous behaviour to classical weighting \cite{Hush2013,*Hush2009,Jacobs2010,Arulampalam2002,*Rebeschini2013}. The mapping is shown schematically in Fig. \ref{fig1}. 

There has been much previous work where ancillas were used to find specific stochastic unravellings for non-Markovian evolution \cite{Breuer2004,Breuer1999,Moodley2009,Imamoglu1994,Stenius1996,Gambetta2002,Yang2012,Stace2006}, or to create non-Markovian evolution in a Markovian system \cite{Budini2013,Budini:2013,Budini:2014a,Siegle2010}. References \cite{Breuer2004,Breuer1999,Budini2013} in particular, served as inspiration, and can now be thought of as special cases in our framework of {\emph{quantum weighting}}.

Our aim is to describe the evolution of $\rhosys$ of the system under a TLME Eq.~(\ref{eqn:genTLME}) in terms of the evolution of $\rhot$, for the same system coupled to an ancilla, under a QME of the form Eq.~(\ref{eqn:introgenMarkov}), where $\rhosys$ is obtained from $\rhot$ via the mapping Eq.~(\ref{eqn:quantweight}): 
\begin{equation}
\dot{\rhot} = \sL_{\rm tot}(\rhot) \rightarrow
\dot \rhosys = \partial_t \Tr_{\rm a}[w_{t}\rhot] = \sK(\rhosys) . \label{primarygoal}
\end{equation}
Note that while the operators in Eq.~(\ref{eqn:genTLME}) act on the system, the self-adjoint $\Ht$ and jump operators $\Jt_{i}$ ($i=1,\ldots,N$) that define the QME Eq.~(\ref{eqn:introgenMarkov}) will act on the \emph{combined} system and ancillary Hilbert space.

We make no additional assumptions about the initial condition of the ancilla and the system $\rhot(0)$, only that it must satisfy Eq.~(\ref{eqn:quantweight}), i.e. $\rhosys(0) = \Tr_{\rm a}[w_{0}\rhot(0)]$. We are not attempting to physically reproduce the dynamics of a non-Markovian bath using our ancillary system. As such, the correlations between the ancilla and the system are not of physical significance in our framework. Consequently, we would expect any issues that arise when mapping non-Markovian time non-local master equations to TLME to still be present once mapped using our approach. In particular, we would not expect the singularities present in the TLME formalism for non-Markovian dynamics to be made regular.


\subsection{Restrictions of Closure} \label{eqn:formofsupG}

Previous work \cite{Breuer2004,Breuer1999} presented specific mappings from Eq.~(\ref{eqn:genTLME}) to a Lindblad master equations which had the property described by Eq.~(\ref{primarygoal}). We also present specific mappings in section \ref{sec:smalancill} (although using a smaller ancilla than in \cite{Breuer2004}). But before looking at specific cases we first develop a generic framework which can be used by the reader to find their own maps tailored to their specific needs.


In Eq.~(\ref{primarygoal}) we posit that the dynamics of the system under the TLME should be recovered by tracing out over the ancilla as in Eq.~(\ref{eqn:quantweight}) with an appropriate choice of weight $w_{t}$, an operator on the ancillary space. Note that Eq.~(\ref{eqn:genTLME}) is \emph{closed} with regard to $\rhosys$, meaning it does not depend on past values or any external free variables. Consequently, the equation of motion we generate for $\rhosys$ must also be \emph{closed}, meaning the RHS of the master equation for $\rho$ must only depend on $\rho$. 

Explicitly this means a valid map will have $\sLtot$ and $w_t$ with the property:
\begin{equation}
\dot{\rhosys} = \partial_t \Tr_{\rm a}[ w_{t} \rhot] \equiv \sM( \Tr_{\rm a}[ w_{t} \rhot]) = \sM(\rhosys). \label{eqn:closureprop} 
\end{equation}
Here $\sM(\rhosys)$ is for the moment an unspecified linear superoperator on $\rhosys$ \emph{only} (not on $\rhot$). We refer to Eq.~(\ref{eqn:closureprop}) as the \emph{closure requirement} \cite{Note1}. Satisfying Eq.~(\ref{eqn:closureprop}) is clearly a weaker, but necessary condition for Eq.~(\ref{primarygoal}) to hold. A general $w_t$ and $\sLtot$ are not guaranteed to satisfy Eq.~(\ref{eqn:closureprop}).

We base our generic framework on this requirement of closure. Although we do not prove what form $\sLtot$ and $w_t$ must have in complete generality, we are able to prove what conditions are put on their form if we apply an ansatz and require closure on each individual term in the master equation (some discussion of weakening this assumption is given in Appendix \ref{apx:lessassumptions}).

We first consider the implications Eq.~(\ref{eqn:closureprop}) has on the weighting operator $w_t$. We take the time derivative of both sides of Eq.~(\ref{eqn:quantweight}),
\begin{align}
\dot{\rhosys} = \Tr_a[\dot{w}_t \rhotot] + \Tr_a[w_t\dot{\rhotot}]. \label{eqn:timeweightevo}
\end{align}
We apply our ansatz and assume the closure condition Eq.~(\ref{eqn:closureprop}) applies to each individual term of Eq.(\ref{eqn:timeweightevo}). When applied to the first term, this results in a restriction on the time evolution of our quantum weighting. In order for the first term in Eq.~(\ref{eqn:timeweightevo}) to be closed, we require $\dot{w}_t = \alpha w_t$, where $\alpha(t)$ is some function, so that 
\be
w_t = e^{ \int_0^t dt' \alpha(t')} w, \label{eqn:weightevo}
\ee
where $w$ is some constant operator that operates on the ancilla. 

In order to understand what the implications the closure property has on the second term of Eq.(\ref{eqn:timeweightevo}) we need to give an explicit form for the combined system-ancilla density matrix evolution:
\begin{align}
\dot {\rhotot} = \sLtot(\rhotot) & = \sLsys(\rhotot) -i\left[\sum_j^{J_c}  M_j \otimes f_j +  M^\dag_j \otimes f_j^\dag ,\rhotot \right] \nn \\ 
& + \sum_k^{K_d} \sD\left[\sum_j^{J_d} N_{j,k}  \otimes g_{j,k} \right], (\rhotot) \label{eqn:genMarkov}
\end{align}
where  $J_c$, $J_d$ and $K_d$ are all naturals, $f_j,g_{j,k}$ are operators on the ancilla, $M_j,N_{j,k}$ are operators on the system only, $\sLsys(\cdot)$ is a Lindblad superoperator which operates on the system only, and $\sD[X](\cdot) \equiv X \cdot X^{\dag} - \frac{1}{2} \{ X^{\dag} X , \cdot \}$. We allow $f_j,g_{j,k},M_j$ and $N_{j,k}$ to be time dependent.  This means the system in some cases will technically be time-dependent Markovian (as discussed in Refs.\ \cite{Breuer2009,Wolf2008}). The inclusion of $\sLsys(\cdot)$ over-specifies the problem slightly, as it could be considered as a special case of the main form where $f_j = g_{j,k} = 1$. But we consider it to be some kind of intrinsic Markovian evolution of the system which cannot be modified, thus it is considered separately. We are primarily interested in engineering the evolution of the ancilla and its coupling to the system. 

We now replace the definition for the time evolution of $w_t$, Eq.~(\ref{eqn:weightevo}), and the explicit form for the Lindblad superoperator $\sLtot$, Eq.(\ref{eqn:genMarkov}), into Eq.~(\ref{eqn:timeweightevo}): 
\begin{align}
\dot{\rhosys} =& \sLsys(\rhosys) + \alpha(t) \rhosys +  e^{\int_0^t ds \; \alpha(s)} \Bigg( \sum_j^{J_c} ( -iM_j \Tr_a[w f_j \rhotot] \nn \\
&-iM^\dag_j \Tr_a[w f^\dag_j \rhotot] +i\Tr_a[f_j w \rhotot] M_j +i \Tr_a[f^\dag_j w \rhotot] M^\dag_j) \nn \\
& + \frac{1}{2}  \sum_{i,j}^{J_d} \sum_{k}^{K_d} ( 2N_{i,k} \Tr_a[ g_{j,k}^\dag w g_{i,k} \rhotot ] N_{j,k}^\dag \nn \\
& -N_{j,k}^\dag  N_{i,k}\Tr_a[ w g_{j,k}^\dag g_{i,k}   \rhotot ] \nn \\ 
&-\Tr_s[ g_{j,k}^\dag g_{i,k} w \rhotot ] N_{j,k}^\dag N_{i,k}) \Bigg),  \label{eqn:wgenMarkov}
\end{align}
where we have suppressed the tensor product notation. Applying our ansatz again, we require each individual term in Eq.~(\ref{eqn:wgenMarkov}) is closed, as described by Eq.~(\ref{eqn:closureprop}). For example the first term enforces $w f_j =  \gamma^{\rm l}_j w$, where $\gamma^{\rm l}_j$ is a complex constant. Applying this ansatz to each term in Eq.~(\ref{eqn:wgenMarkov}) gives:
\begin{subequations} \label{eqn:preweightalgebra}
\begin{align}
w f_j = \gamma^{\rm l}_j w; \; & w f^\dag_j  =  \mu^{\rm l}_j w,\label{eqn:preweightalgebraa}   \\
 f_j w  = \gamma^{\rm r}_j w; \; &  f_j^\dag w  = \mu^{\rm r}_j w,\label{eqn:preweightalgebrad}   \\
 g_{j,k}^\dag w g_{i,k} & = \kappa_{i,j,k}^{\rm m} w,\label{eqn:preweightalgebrac} \\
 w g_{j,k}^\dag g_{i,k} & = \kappa_{i,j,k}^{\rm l} w, \label{eqn:preweightalgebrab} \\
 g_{j,k}^\dag g_{i,k}  w &= \kappa_{i,j,k}^{\rm r} w. \label{eqn:preweightalgebrae}
\end{align}
\end{subequations}
Here $\gamma^{\rm l}_j,\gamma^{\rm r}_j, \mu^{\rm l}_j,\mu^{\rm r}_j, \kappa_{i,j,k}^{\rm l}, \kappa_{i,j,k}^{\rm m}$ and $\kappa_{i,j,k}^{\rm r}$ are all complex constants \footnote{To minimise confusion we continue to use Roman letters for operators: capitals for operators that act on the system and lower case for operators that act on ancilla. We also add an additional convention of using Greek symbols for constants or vectors. There are two exceptions to this convention: the density matrix is expressed as $\rho$ or $
\varrho$; and sums are labelled with Roman letters, but are always either $i$, $j$, $k$ or $l$ (caps or no caps). }. 

We have transformed the abstract closure property defined in Eq.~(\ref{eqn:closureprop}) into an explicit set of equations the weighting and coupling operators must obey: Eqs.~(\ref{eqn:preweightalgebra}). Our ansatz, namely each individual term of Eq.~(\ref{eqn:wgenMarkov}) must satisfy the closure condition Eq.~(\ref{eqn:closureprop}), is a strong requirement. However, it can be weakened and this is discussed further in Appendix \ref{apx:lessassumptions}. But to simplify working, in what follows we take the ansatz for granted. We can now state the central theorem of our framework:

\emph{Theorem:} Given a system and ancilla whose evolution is governed by Eq.~(\ref{eqn:genMarkov}), if the coupling and weighting operators satisfy Eq.~(\ref{eqn:weightevo}) and Eqs.~(\ref{eqn:preweightalgebra}) (which in turn satisfies the closure property, Eq.~(\ref{eqn:closureprop})) then the linear superoperator $\sM(\rhosys)$ (from Eq.~(\ref{eqn:closureprop})) must have the form:
\begin{align}
\sM(\rhosys) & = \sLsys(\rhosys) +\alpha \rhosys + \left(A_l - S_l  - \frac{1}{2}\sum_{i=1}^I L_i^\dag L_i \right) \rhosys \nn \\
& + \rhosys \left(A_r - S_r - \frac{1}{2} \sum_{i=1}^I R_i^\dag R_i \right)  + \sum_{i=1}^I  L_i \rhosys R_i^\dag, \label{eqn:effnotMarkovevo}
\end{align}
where $I$ is a natural, $A_{r/l}$ are anti-hermitian operators, $S_{r/l}$ are positive-semidefinite operators, and $L_i$ and $R_i$ general operators.

Eq.~(\ref{eqn:effnotMarkovevo}) is the central relation of the paper. It holds for all mappings that obey our restricted notion of closure. In section \ref{sec:genmap} we will show that even with this restriction it is always possible to change the form of $\sM(\rhosys)$ into $\sK(\rhosys)$. However, in section \ref{sec:physrel} we will show the restrictions on $\sM(\rhosys)$ still have important implications on the norm growth that can make sampling in an experimental Markovian implementation of Eq.~(\ref{eqn:genTLME}) inefficient. 

We first prove a lemma that determines what restrictions Eqs.~(\ref{eqn:preweightalgebraa}) puts on the operators in Eq.~(\ref{eqn:genMarkov}). These restrictions are presented as a set of algebraic equations.

\emph{Lemma:} If we require Eq.~(\ref{eqn:preweightalgebra}) all hold \emph{simultaneously} then:
\begin{subequations} \label{eqn:constrestr}
\begin{align}
\mu_j^{\rm l} = & (\gamma_j^{\rm l})^*,  \label{eqn:constrestra} \\
\mu_j^{\rm r} = & (\gamma_j^{\rm r})^*,  \label{eqn:constrestre} \\
\kappa_{i,j,k}^{\rm m} = & \inpr{ \delta_{j,k}^{\rm r} }{ \delta_{i,k}^{\rm l} }, \label{eqn:constrestrc} \\
\kappa_{i,j,k}^{\rm l} = &\inpr{ \delta_{j,k}^{\rm l} }{ \delta_{i,k}^{\rm l} } + \inpr{ \epsilon_{j,k}^{\rm l} }{ \epsilon_{i,k}^{\rm l} }, \label{eqn:constrestrd} \\
\kappa_{i,j,k}^{\rm r} = & \inpr{ \delta_{j,k}^{\rm r} }{ \delta_{i,k}^{\rm r} } + \inpr{ \epsilon_{j,k}^{\rm r} }{ \epsilon_{i,k}^{\rm r} }, \label{eqn:constrestrb}
\end{align}
\end{subequations}
where $\delta$ and $\epsilon$ are vectors and $( \cdot | \cdot)$ is our notation for an inner product.

\emph{Proof:} First consider the equations: $w f_j = \gamma^{\rm l}_j w$ and $w f^\dag_j  =  \mu^{\rm l}_j w$ from Eq.~(\ref{eqn:preweightalgebraa}). We apply the Moore-Penrose pseudoinverse \cite{,Penrose1955} of $w$, $w^+$, to the LHS of both equations:
\begin{subequations} 
\begin{align}
p_l f_j = \gamma^{\rm l}_j p_l; \label{eqn:lhspsincohcondA} \\
p_l f^\dag_j  =  \mu^{\rm l}_j p_l. \label{eqn:lhspsincohcondB}
\end{align}
\end{subequations}
Where $p_l=w^+w$ is a projector with $p_l=0$ only when $w=0$ (which we have assumed is not the case). Using Eq.~(\ref{eqn:lhspsincohcondA}) we can show $p_l f_j p_l = \gamma^{\rm l}_j p_l^2 = \gamma^{\rm l}_j p_l$ and using Eq.~(\ref{eqn:lhspsincohcondB}) we can show $p_l f^\dag_j p_l =  \mu^{\rm l}_j p_l^2 = \mu^{\rm l}_j p_l$, as $p_l f_j p_l = (p_l f^\dag_j p_l)^\dag$ this implies:
\be
\gamma^{\rm l}_j = (\mu^{\rm l}_j)^*.  \label{eqn:constrestraresulta}
\ee
Eq.~(\ref{eqn:constrestraresulta}) proves  Eq.~(\ref{eqn:constrestra}).

Using equations: $f_j w  = \gamma^{\rm r}_j w$ and $f_j w  = \mu^{\rm r}_j w$ from Eq.~(\ref{eqn:preweightalgebrad}) it can be shown 
\be
\gamma^{\rm r}_j = (\mu^{\rm r}_j)^*, \label{eqn:constrestraresultb}
\ee
with the same methodology used to prove Eq~(\ref{eqn:constrestraresulta}). This proves Eq.~(\ref{eqn:constrestre}).

Proving Eq.~(\ref{eqn:constrestrc}) can be achieved using singular value decomposition (SVD) \cite{Horn1985}. We use SVD to factorise $w$ into $w= u s v^\dag$ where $u$ and $v$ are unitary matrices and $s$ is a diagonal matrix containing the singular values of $w$. The singular values are all strictly positive and the rank of $s$ is guaranteed to be greater than 1 for $w \ne 0$.  Replacing this decomposition into $g_{j,k}^\dag w g_{i,k} = \kappa_{i,j,k}^{\rm m} w$, from Eq.~(\ref{eqn:preweightalgebrac}), gives $g_{j,k}^\dag usv^\dag g_{i,k} = \kappa_{i,j,k}^{\rm m} usv^\dag$, which we can rearrange into: 
\be
p u^\dag g_{j,k}^\dag u s v^\dag g_{i,k} v s^+  = \kappa_{i,j,k}^{\rm m} p, \label{eqn:projectorresultm}
\ee
where $p = s s^+$ is a projector. We can express Eq.~(\ref{eqn:projectorresultm}) in terms of the elements of the matrix as follows
\be
\sum_{l}^P \frac{\{\sqrt{s}u^\dag g_{j,k} u\}_{l,m}^* \{\sqrt{s} v^\dag g_{i,k} v \}_{l,n}}{\{s\}_{n,n}}  = \kappa_{i,j,k}^{\rm m} \delta_{m,n}. \label{eqn:elementsresultsm}
\ee
where $\{ \cdot\}_{m,n}$ is our notation for taking the $(m,n)$ element of a matrix and the matrix indices go from 1 to the rank of $p$ which we define as $P$. Eq.~(\ref{eqn:elementsresultsm}) is a set of equations which must be satisfied for $w$ and $g_{i,k}$. We assume an appropriate set of matrices have already been found as we are only interested in putting restrictions on $\kappa_{i,j,k}^{\rm l/r/m}$. For this we consider the equation in Eq.~(\ref{eqn:elementsresultsm}) corresponding to the maximum $s$, i.e. we take $m=n=M$ where $\{s\}_{M,M} = \mbox{max}[s]\equiv s_{\rm max}$, this equation is equivalent to: 
\be 
\kappa_{i,j,k}^{\rm m} = \inpr{ \delta^{\rm r}_{j,k} }{ \delta^{\rm l}_{i,k} }, \label{proofofkappam}
\ee
where  $\{\delta^{\rm l}_{i,k}\}_l \equiv \{\sqrt{s/s_{\rm max}} v^\dag g_{i,k} v\}_{l,M} $ and $\{\delta^{\rm r}_{j,k}\}_l \equiv \{ \sqrt{s/s_{\rm max}} u^\dag g_{j,k} u \}_{l,M} $ are vectors and $ \inpr{ \cdot  }{ \cdot } = \sum_l \{ \cdot \}_l^* \{ \cdot \}_l$.

We move our attention to Eq.~(\ref{eqn:constrestrd}). We start by replacing the SVD of $w=usv^\dag$ into $w g_{j,k}^\dag g_{i,k} = \kappa_{i,j,k}^{\rm l} w$ from Eq.~(\ref{eqn:preweightalgebrab}), after rearranging we find:
\be 
p v^\dag g_{j,k}^\dag v \left( \frac{s}{s_{\rm max}} + p - \frac{s}{s_{\rm max}} + q\right) v^\dag g_{i,k} v p  = \kappa_{i,j,k}^{\rm l} p. \label{eqn:projectorresultl}
\ee
Where we defined $q = 1 -p $ which is another projector, and used the identity $1 = s/s_{\rm max} + p - s/s_{\rm max} + q$. Again if we looked at the matrix elements of Eq.~(\ref{eqn:projectorresultl}), we would have an set of equations that have to be satisfied. If we take the same approach used to create Eq.~(\ref{proofofkappam}),  and take the equation corresponding to matrix element $m=n=M$, we find:
\be
\kappa_{i,j,k}^{\rm l} =\inpr{ \delta_{j,k}^{\rm l} }{ \delta_{i,k}^{\rm l} } + \inpr{ \epsilon_{j,k}^{\rm l} }{ \epsilon_{i,k}^{\rm l} } \label{eqn:proofofkappal}
\ee 
Where $\delta_{i,k}^{\rm l}$ was previously defined and $\{ \epsilon_{i,k}^{\rm l} \}_l =  \{ (\sqrt{p - s/s_{\rm max}} + q) v^\dag g_{i,k} v p \}_{l,M}$ and $l$ in this case actually goes from 1 to the rank of $p+q$ (instead of only up to the rank of $p$). Note that $\sqrt{p - s/s_{\rm max} } \ge0$ by definition, this ensures we can express the term $\inpr{ \epsilon_{j,k}^{\rm l} }{ \epsilon_{i,k}^{\rm l} }$ as an inner product. This proves Eq.~(\ref{eqn:constrestrd}). 

There is currently a complicated interdependence between the vectors $\epsilon_{i,j}^{\rm l}$ and $\delta_{i,k}^{\rm l}$. Fortunately, this will not have any impact on the rest of this proof. If one was engineering a particular ancilla-system coupling and found this interdependence to be a problem, one can always select a $w$ which has equal singular values, e.g. $s=p$, in which case $\epsilon_{i,k}^{\rm l}$ becomes independent from $\delta_{i,k}^{\rm l}$. 

Lastly we consider Eq.~(\ref{eqn:constrestrb}), starting with $g_{j,k}^\dag g_{i,k} w  = \kappa_{i,j,k}^{\rm r} w$ from Eq.~(\ref{eqn:preweightalgebrab}) and applying the same procedure used to derive Eq.~(\ref{eqn:proofofkappal}) it can be shown:
\be
\kappa_{i,j,k}^{\rm r} = \inpr{ \delta_{j,k}^{\rm r} }{ \delta_{i,k}^{\rm r} } + \inpr{ \epsilon_{j,k}^{\rm r} }{ \epsilon_{i,k}^{\rm r} } \label{eqn:proofofkappar}
\ee
where $\delta_{i,k}^{\rm r}$ was previously defined and $\{\epsilon_{i,k}^{\rm r}\}_l =  \{(\sqrt{p - s/s_{\rm max}} + q) u^\dag g_{i,k} u \}_{l,M}$. This proves Eq.~(\ref{eqn:constrestrb}), which completes the proof of Eq.~(\ref{eqn:constrestr}) \qed

We now prove the main theorem. We replace Eq.~(\ref{eqn:constrestr}) into Eq.~(\ref{eqn:preweightalgebra}), to get what we term the \emph{quantum weight algebra}. This is the algebra both the ancillary coupling operators and weighting operator must obey:
\begin{subequations} \label{eqn:ginalweightalgebra} 
\begin{align}
w f_j &  = \gamma^{\rm l}_j w; \; w f_j^\dag  = (\gamma^{\rm l}_j)^* w; \; \\ 
f_j w  & = \gamma_j^{\rm r} w; \; f_j^\dag w  = (\gamma^{\rm r}_j)^* w;  \\
g_{j,k}^\dag w g_{i,k} & = \inpr{ \delta_{j,k}^{\rm r} }{ \delta_{i,k}^{\rm l} } w; \\
w g_{j,k}^\dag g_{i,k} & = (  \inpr{ \delta_{j,k}^{\rm l} }{ \delta_{i,k}^{\rm l} } + \inpr{ \epsilon_{j,k}^{\rm l} }{ \epsilon_{i,k}^{\rm l} } )w; \\
{\rm and} \; g_{j,k}^\dag g_{i,k}  w& = (\inpr{ \delta_{j,k}^{\rm r} }{ \delta_{i,k}^{\rm r} } + \inpr{ \epsilon_{j,k}^{\rm r} }{ \epsilon_{i,k}^{\rm r} }) w.
\end{align}
\end{subequations}
Where the constants were previously defined. 

We now assume that some set of operators for $w$, $f_j$ and $g_{i,k}$ have been found that obey Eq.~(\ref{eqn:ginalweightalgebra}). Using these operators we can find the form of $\sM(\rhosys)$ and determine if the closure property has resulted in it being restricted. We achieve this goal by replacing Eq.~(\ref{eqn:ginalweightalgebra}) into Eq.~(\ref{eqn:wgenMarkov}):
\begin{align}
\dot{\rhosys} & = \sM(\rhosys) \equiv \sLsys(\rhosys)+ \alpha(t) \rhosys \nn \\
&+ \sum_j^{J_c} (-i (\gamma_j^{\rm l} M_j + (\gamma_j^{\rm l})^* M^\dag_j) \rhosys + i \rhosys( \gamma_j^{\rm r} M_j + (\gamma_j^{\rm r})^* M^\dag_j) ) \nn \\
& + \sum_{i,j}^{J_d} \sum_{k}^{K_d} \Big( N_{i,k} \rhosys N_{j,k}^\dag \inpr{ \delta_{j,k}^{\rm r} }{ \delta_{i,k}^{\rm l} }  \nn \\ 
&- \frac{1}{2} ( \inpr{ \delta_{j,k}^{\rm l} }{ \delta_{i,k}^{\rm l} } + \inpr{ \epsilon_{j,k}^{\rm l} }{ \epsilon_{i,k}^{\rm l} } ) N_{j,k}^\dag N_{i,k} \rhosys  \nn \\
&- \frac{1}{2}  ( \inpr{ \delta_{j,k}^{\rm r} }{ \delta_{i,k}^{\rm r} } + \inpr{ \epsilon_{j,k}^{\rm r} }{ \epsilon_{i,k}^{\rm r} } ) \rhosys N_{j,k}^\dag N_{i,k} \Big), \label{eqn:compnonMarkov}
\end{align}
We can simplify the form of this equation by taking advantage of our inner product notation. If we define $| \cdot )^\dag = ( \cdot|$ and use $(b|a)= ( a^*| b^*)$ Eq.~(\ref{eqn:compnonMarkov}) reduces to 
\begin{align}
\dot{\rhosys} & = \sM(\rhosys) \equiv \sLsys(\rhosys)+ \alpha(t) \rhosys \nn \\
&+ \sum_j^{J_c} (-i (\gamma_j^{\rm l} M_j + (\gamma_j^{\rm l})^* M^\dag_j) \rhosys + i \rhosys( \gamma_j^{\rm r} M_j + (\gamma_j^{\rm r})^* M^\dag_j) ) \nn \\
& + \sum_{k}^{K_d} \Big( N_{k}^{\delta, \rm l} \rhosys (N_{k}^{\delta,\rm r})^\dag - \frac{1}{2} ( (N_{k}^{\delta,\rm l})^\dag N_{k}^{\delta, \rm l} \rhosys + \rhosys (N_{k}^{\delta,\rm r})^\dag N_{k}^{\delta,\rm r} ) \nn \\
&  - \frac{1}{2} ( (N_{k}^{\epsilon, \rm l})^\dag N_{k}^{\epsilon,\rm l} \rhosys + \rhosys (N_{k}^{\epsilon,\rm r})^\dag N_{k}^{\epsilon,\rm r} ) \Big), \label{eqn:fincompnonMarkov}
\end{align}
where $N_k^{\delta/\epsilon,\rm l/r} = \sum_i^{J_d} N_{i,k} | (\delta/\epsilon)^{\rm l/r}_{i,k} )$. The operators $N_k^{\delta/\epsilon,\rm l/r}$ for a general ancilla-system coupling have a complicated interdependence. However, a particular ancilla-system coupling can always be found that allows us to treat them all as independent operators which act on only on the system space. This is discussed in more detail in appendix \ref{apx:complicatedNepdellr}. Whether or not such an ancilla-system coupling is being used does not affect the rest of this proof, so we continue without making any such assumptions.

We have technically achieved our primary goal of finding the form of $\sM(\rhosys)$, Eq.~(\ref{eqn:fincompnonMarkov}). However, if one was given a superoperator it would be very difficult to state if it is in the form of Eq.~(\ref{eqn:fincompnonMarkov}) or not. We have not provided a clear understanding on what restrictions the closure property puts on $\sM(\rhosys)$. In order to make this more transparent we identify specific features the operators must possess to be in the form of  Eq.~(\ref{eqn:fincompnonMarkov}).

We recognise: $-i (\gamma_j^k M_j + (\gamma_j^k)^* M^\dag_j) \rhosys$, the second term in Eq.~(\ref{eqn:compnonMarkov}), can only generate terms which operate on the LHS of $\rhosys$ and are anti-Hermitian; next, $ - \frac{1}{2}\rhosys (N_k^{\delta,r})^\dag N_k^{\delta,r}$ can only generate terms which operate on the RHS of $\rhosys$ and are negative semidefinite. Systematically applying this logic we get the following form: 
\begin{align}
\sM(\rhosys) & = \sLsys(\rhosys) +\alpha(t) \rhosys + (A_l - S_l) \rhosys + \rhosys (A_r - S_r) \nn \\
&  + \sum_i^I  L_i \rhosys R_i^\dag -\frac{1}{2} (L_i)^\dag L_i \rhosys-\frac{1}{2} \rhosys (R_i)^\dag R_i, \label{eqn:effnotMarkovevofinal}
\end{align}
where $A_l$ and $A_r$ are arbitrary anti hermitian operators, $S_l$ and $S_r$ are positive-semidefinite operators and $L_l$ and $L_r$ are arbitrary operators. 

To make this connection clearer, we give the explicit relationship between Eqs.~(\ref{eqn:fincompnonMarkov}) and (\ref{eqn:effnotMarkovevofinal}):
\begin{subequations} \label{eqn:ASLRtoKMN}
\begin{align}
A_{\rm l/r} =& \sum_j^{J_c} (-i (\gamma_j^{\rm l/r} M_j + (\gamma_j^{\rm l/r})^* M^\dag_j); \\
S_{\rm l/r} =& \frac{1}{2} \sum_{k \in \bK_d^{\rm l/r}} (N_k^{\delta, \rm l/r})^\dag N_k^{\delta, \rm l/r} \nn \\
& + \frac{1}{2}  \sum_{k}^{K_d}(N_k^{\epsilon, \rm l/r})^\dag N_k^{\epsilon, \rm l/r}; \label{eqn:semiposdefn} \\ 
L_i =& N^{\rm l}_i \; \forall \; i\in \bK_d^{\rm m}; \label{eqn:leftdefn} \\
R_i =& N^{\rm r}_i \; \forall \; i\in \bK_d^{\rm m}. \label{eqn:righdefn} 
\end{align}
\end{subequations}
where we grouped the $N_k^{\delta,\rm l/r}$ operators as follows: $\bK_d^{\rm l} = \{k: N_k^{\delta,\rm r}=0\}$, $\bK_d^{\rm r} =\{k:N_k^{\delta,l} = 0\}$ and $\bK_d^{\rm m} = \{k:N_k^{\delta,\rm l/r} \ne 0\}$.

Operators $A_{\rm l/r}$ and $S_{\rm l/r}$, defined in Eq.~(\ref{eqn:ASLRtoKMN}), are clearly anti-Hermitian and positive-semidefinite respectively. This completes the theorem as Eq.~(\ref{eqn:effnotMarkovevofinal}) is precisely in the form of Eq.~(\ref{eqn:effnotMarkovevo}) as required. \qed

For a general ancilla-system coupling there may be a complicated interdependence between $L_i/R_i$ and $S_{\rm l/r}$ given in Eq~(\ref{eqn:effnotMarkovevofinal}). However, such interdependence will not change the form of $\sM(\rhosys)$. Also,  a particular ancilla-system coupling can always be picked such that $L_i/R_i$ and $S_{\rm l/r }$ are all independent operators. This is further discussed in appendix \ref{apx:complicatedNepdellr}.

\subsection{Generic Mapping} \label{sec:genmap}

We have determined the algebraic form of $\sM(\rhosys)$, Eq.~(\ref{eqn:effnotMarkovevo}) although it  looks restrictive we can actually carefully select the function $\alpha(t)$ and operators $A_{\rm l/r}$, $S_{\rm l/r}$, $L_i$ and $R_i$ such that $\sM(\rhosys)=\sK(\rhosys)$. In section \ref{sec:physrel} we will see that a large $\alpha(t)$ can result in a sampling inefficiency. Thus we pick the operators such that $\alpha(t)$ is minimised. 

First we identify 
\be 
L_i = D_i  \mbox{ and } R_i = E_i \label{eqn:LRtoDE}
\ee
(and so $I=M$).  Second, we identify $A_{\rm r,l}$ with the anti-Hermitian parts of $B,C$, that is, 
\be
A_{\rm l} = B_- \mbox{ and } A_{\rm r} = C_- \label{eqn:AtoBC}
\ee
where $X_{\pm} \equiv (X \pm X^\dag)/2$.  Third, we note the Hermitian parts of $B$ and $C$ are not necessarily negative semidefinite (which the form of $\sM(\rhosys)$, Eq.~(\ref{eqn:effnotMarkovevo}), requires), but can always be made so by the subtraction of a sufficiently large constant, so we define 
\begin{equation}
S_{\rm l} \equiv \alpha_{\rm l} - H_{ \rm l}; \; {\rm and} \; S_{\rm r} \equiv \alpha_{\rm r} - H_{ \rm r}. \label{defS} 
\end{equation}
Where $H_{\rm l} = B_{+} + \sum_i^M D_i^\dag D_i/2$ and $H_{\rm r} = C_{+} + \sum_i^M E_i^\dag E_i/2$ and the constants $\alpha_{\rm r,l}$ are defined by
\begin{equation}
\alpha_{\rm l} \equiv \lambda_{\rm max}^{\rm +}\left[ H_l \right]; \; {\rm and} \; \alpha_{\rm r} \equiv \lambda_{\rm max}^{\rm +}\left[ H_r \right]. \label{alphalr}
\end{equation}
Where $\lambda_{\rm max}^{\rm +}(X)$ returns the largest positive eigenvalue of matrix $X$, or returns $0$ if $X$ has no positive eigenvalues. The operators $H_{\rm l/r}$, and consequently $\alpha_{\rm l/r}$, can be time dependent, but this does not affect the central result.  Fourth, we are forced to set 
\be
\alpha = \alpha_{\rm l} + \alpha_{\rm r}, \label{eqn:minalphadef}
\ee
as this is only way the negative constants $-\alpha_{\rm l/r}\le 0$ can be cancelled according to the form of $\sM(\cdot)$, Eq.~(\ref{eqn:effnotMarkovevo}). With these identifications Eq.~(\ref{eqn:effnotMarkovevo}) becomes of the form Eq.~(\ref{eqn:genTLME}), guaranteeing that any TLME can be obtained from a QME in an enlarged Hilbert space. Note that $\alpha_{\rm r,l}$ are the smallest possible numbers that could be added to the operators $S_{\rm l}$, $S_{\rm r}$ to ensure they are positive semidefinite. Consequently, $\alpha$ is minimised as required. Note: $\alpha$, physically, is the norm \emph{growth} of Eq.~(\ref{eqn:genTLME}).

We have found a generic mapping between a TLME~(\ref{eqn:genTLME}) and a Lindblad master equation~(\ref{eqn:introgenMarkov}). The map is underspecified, meaning for a given TLME, there is a multitude of equivalent Lindblad master equations. Nonetheless, using the equations derived finding a map is no longer a process of guess and check instead it is simplified to solving a set of indeterminate matrix equations. The algorithm goes as follows: 
\begin{enumerate}
\item Change the TLME into the form of Eq.~(\ref{eqn:genTLME}).
\item Use Eqs.~(\ref{eqn:LRtoDE}) - (\ref{eqn:minalphadef}) to get $\alpha(t)$, $A_{\rm l/r}$, $S_{\rm l/r}$, $L_i$ and $R_i$ in terms of $A$, $B$, $C$ and $D$.
\item Simultaneously solve the underspecified matrix equations Eqs.~(\ref{eqn:ASLRtoKMN}) and (\ref{eqn:ginalweightalgebra}) to find $ w$,  $f_j$, $g_{j,k}$, $M_j$ and $N_{j,k}$.
\item Replace selected solution into Eq.~(\ref{eqn:genMarkov}) to get a Markovian master equation in form of Eq.~(\ref{eqn:introgenMarkov}) as required. Moments for the TLME, Eq.~(\ref{eqn:genTLME}), can be calculated using the quantum weighted average Eq.~(\ref{eqn:quantweight}) and the solution for $w_t$, Eq.~(\ref{eqn:weightevo}).
\end{enumerate}

The mapping the reader chooses will depend on their precise application. We expect that in most cases, the smaller the better. Hence, we have used the algorithm above to find: a map between a generic TLME and a Markovian system with the \emph{smallest possible} ancilla.

\subsection{Smallest possible ancilla}  \label{sec:smalancill}

Now that we have a method for deriving a map between TLME and Markovian equations, a basic question is then: how big must the ancillary system be? As it happens, the minimum necessary ancillary Hilbert space is that of a qubit. We show this by providing two explicit general mappings.  We denote the basis states of the ancilla by $|0\rangle$ and $|1\rangle$, and consider two possible choices of the weight operator $w$ (and from which other choices can be deduced): (i) $w$ is diagonal, say $w = |0\rangle \langle 0|$; (ii) is off-diagonal, say $w = |1\rangle \langle 0|$.

\noindent
{\em (i) Diagonal ancilla and Hermiticity-preserving TLMEs:} A mapping with $w = |0\rangle \langle 0|$ is appealing because in this case the weight operator $w_t$ is Hermitian at all times, $w_t=w_t^\dag$, and is therefore an observable.  This mapping is only possible for TLMEs which maintain the Hermiticity of the system density matrix $\rhosys$ (given that the system-plus-ancilla $\rhot$ is Hermitian at all times).  This means that we can simulate all TLMEs where $C \equiv B^\dag$ and $D_i \equiv E_i$.  In this case, the following QME for the system-plus-ancilla gives the TLME evolution of the system under the mapping Eq.~(\ref{eqn:quantweight}),
\begin{align}
\dot{\rhot} =& \sL_{\rm sys}(\rhot) + [ B_- \otimes 1 ,\rhot] 
+ \sD[ \sqrt{2 S_{\rm l}} \otimes \sigma](\rhot)
\nn \\
& 
+ \sum_{i=1}^M \sD[ D_i \otimes 1 ](\rhot) . 
\label{eqn:nonHermMarkovevo}
\end{align}
Here $\sL_{\rm sys}$ acts only on the system and not on the ancilla, $\sigma \equiv |1\rangle \langle 0|$ and $S_{\rm l}$ still has the same definition as Eq.~(\ref{defS}). The TLME can be recovered from Eq.~(\ref{eqn:nonHermMarkovevo}) using the quantum weighting: $\rho = \Tr_a[w_t \rhot]$, Eq.~(\ref{eqn:quantweight}).

\noindent
{\em (ii) Off-diagonal ancilla and generic TLMEs:} If $w = |1\rangle \langle 0|$ then the Hermiticity-preserving restrictions above do not apply and a mapping for any TLME can be found,
\begin{align}
\dot{\rhot}& = \sL_{\rm sys}(\rhot) + [ B_- \otimes p_0 + C_- \otimes p_1 ,\rhot] + \sD[\sqrt{S_{\rm l}} \otimes p_0 ] (\rhot)
\nn \\ 
&
+ \sD[\sqrt{S_{\rm r}} \otimes p_1 ] (\rhot) + \sum_{i=1}^M \sD[D_{i} \otimes p_0 + E_{i} \otimes p_1](\rhot)
\label{eqn:notMarkovevo}
\end{align}
where $\sL_{\rm sys}$ only acts on the system, $p_0 \equiv |0\rangle \langle 0|$, $p_1 \equiv |1\rangle \langle 1|$, and the operators $S_{\rm l,r}$ are defined in Eq.~(\ref{defS}). Once again the TLME for $\rho$, Eq.~(\ref{eqn:genTLME}), can be recovered using the quantum weighting as previously defined.

\subsection{ Multiplicity of mappings} \label{sec:guagerel}

The mappings Eqs.~(\ref{eqn:nonHermMarkovevo}) and (\ref{eqn:notMarkovevo}) are not unique. When choosing a mapping, the problem is always underspecified. The question then arises how are these different mappings related? Given a Lindbladian $\sL$ which maps to $\sK$ under $w$, Eq.~(\ref{primarygoal}), one can obtain a second $\sL' = \sL + \sG_t$ which also obeys Eq.~(\ref{primarygoal}), as long as $\Tr_{\rm a}[w_{t} \sG_t(\cdot)] = 0$.  This is a ``gauge invariance'' of the mapping, since we are describing one system by means of a larger one, and so we can make (possibly time-dependent) transformations of the operators on the system-plus-ancilla while maintaining the same dynamics of the system.  This freedom can be exploited to obtain to the most convenient mapping for the problem at hand, as we discuss in Application I.

\subsection{ Sampling inefficiency}  \label{sec:physrel}

A map between a TLME and a Lindblad master equation allows the machinery developed for working with Markovian systems to be applied to TLME. In principal, we can even physically realise TLME with Markovian quantum hardware. However, in an experiment one does not have direct access to the density matrix and instead has to infer information by making measurements of observables. The question arises, assuming a physical realisation has been found, can observables be sampled efficiency?  Here we show when $\alpha(t)>0$ the sampling of observables become inefficient. Specifically the number of measurements required to keep the same precision will grow exponentially with time.

Consider measuring some system observable $X$, 
\begin{equation}
\langle X \rangle = \Tr_{\rm s} [ X~\Tr_{\rm a}[w_t \rhot] ] = e^{\int_0^tdt'\; \alpha(t')} ~ \Tr_{\rm s} [ X~\Tr_{\rm a}[w \rhot] ] .
\nn
\end{equation}
To determine $ \langle X \rangle$ in experiment many repeated measurements will be required to first determine $\Tr_{\rm s} [ X~\Tr_{\rm a}[w \rhot] ]$. In practice, $\Tr_{\rm s} [ X~\Tr_{\rm a}[w \rhot] ]$ will only be known up to some precision.  When $\langle X \rangle$ is calculated this lack of precision will get exponentially amplified in time due to the factor $e^{\int_0^tdt'\; \alpha(t')}$ when $\alpha(t) >0$.  When $\alpha = 0$  this exponential amplification of the uncertainty does not occur. 

This observation allows us to split TLMEs into two distinct classes, depending on whether $\alpha=0$ or $\alpha>0$. When $\alpha > 0$ the uncertainty (or lack of precision) in $\langle X \rangle$ will grow exponentially in time. This means an exponential number of measurements will be required to achieve the same precision for any observable of the system over time.  Thus we state that if $\alpha > 0$ for a given TLME our framework can not be used to find a mapping that can be sampled efficiently.
In contrast, if a TLME has $\alpha(t) = 0$ or $\alpha(t)>0$ for only a finite time it is at least possible that $\langle X \rangle$ can be sampled efficiently.

Lastly we emphasise that this classification of efficient or inefficient TLME can be performed \emph{without finding an explicit mapping to a Markovian system}. One only has to get the TLME in the form of Eq.~(\ref{eqn:genTLME}) (repeated here):
\begin{equation}
\dot{\rho} = \sLsys(\rho) + B\rho + \rho C + \sum_{j=1}^{I} D_j \rho E_j^\dag 
\end{equation}
Then  $\alpha\equiv \alpha(t)$ is defined as Eq.~(\ref{eqn:minalphadef}):
\begin{align}
\alpha = &\lambda_{\rm max}^+\left[\frac{1}{2} (B+B^\dag) + \frac{1}{2} \sum_i^{I} D_i^\dag D_i \right] \nn \\
& + \lambda_{\rm max}^+ \left[\frac{1}{2}(C+C^\dag) + \frac{1}{2} \sum_i^{I} E_i^\dag E_i \right], \label{eqn:repofalpha}
\end{align}
where $\lambda_{\rm max}^+[X]$ returns the largest positive eigenvalue of matrix X or return 0 is X has no positive eigenvalues. If $\alpha(t)=0$ indefinitely or $\alpha(t) >0$ for a finite time an efficient mapping may exist, or if $\alpha(t) >0$ indefinitely no efficient mapping exists. As $\alpha(t)$ is directly related to the norm growth of Eq.~(\ref{eqn:genTLME}), physically we can interpret the inefficiency as being a consequence of norm growth. If an efficient mapping may exist, one may then use the two level ancilla mapping presented section \ref{sec:smalancill} or create a new map using the algorithm from section \ref{sec:genmap} or the gauge freedom described in section \ref{sec:guagerel}.

In summary up to now: we have developed a generic framework for deriving maps between TLME and Lindblad master equations, and presented two maps between a general TLME and Markovian system with an attached two level ancillary system.  Furthermore, we have determined a way of classifying which TLMEs cannot, using our framework, be efficiently sampled when realised in an experiment.  In the next two sections we apply these results in two contexts, first in the thermodynamics of trajectories and second in quantum control.

\section{ Application I: thermodynamics of trajectories} 

Consider a system evolving according to a QME with Lindbladian $\sL_{\rm sys}$ for which we wish to compute the probability $P_{t}(K)$ of observing $K$ quantum jumps due to jump operator, say, $J_{1}$.  Such time-integrated quantities are convenient order parameters for classifying the {\em dynamical} phase structure of open systems \cite{Lecomte2007,Garrahan2010}.  Instead of the probability $P_{t}(K)$ we may consider the generating function $Z_{t}(s) \equiv \sum_{K} e^{-s K} P_{t}(K)$. At large times this acquires a large-deviation (LD) form \cite{Touchette2009}, $Z_{t}(s) \sim e^{t \theta(s)}$, where the LD function (scaled cumulant generating function) $\theta(s)$ plays the role of a free-energy density for trajectories \cite{Lecomte2007,Garrahan2010}, where the ``counting'' field $s$ is conjugate to the observable $K$.  This leads to the definition of the deformed (or ``tilted'') operator $\sW_{s}$ \cite{Garrahan2010},
\begin{equation}
\dot \rhosys =
\sW_{s}(\rhosys)
\equiv
\sL_{\rm sys}(\rhosys) + e^{-s} J_{1} \rhosys J_{1}^{\dag} - \frac{1}{2} \{ J_{1}^{\dag} J_{1} \rhosys \} .
\label{eqn:dynpress}
\end{equation}
 The LD function $\theta(s)$ is given by the largest eigenvalue of $\sW_{s}$, such that, 
\begin{equation}
 \Tr[\rhosys] \sim e^{t \theta(s)} ,
\label{Trs}
\end{equation}
at large times.  The above is a TLME for the evolution of $\rhosys$.  The dynamics it generates is related to that of a subset of trajectories of the original dynamics, reweighed such that the average $K$ is given by $-\theta'(s)$ [and not $-\theta'(0)$ as in the original dynamics] (sometimes called the $s$-ensemble of the dynamics \cite{Hedges2009}).

The general mapping allows to access this $s$-ensemble through the actual dynamics of a system-plus-ancilla.  Since $\sW_{s}$ is Hermiticity-preserving, we can choose case (i) or (ii) for the ancilla.  If we choose $w = |0\rangle \langle 0|$, then from Eq.~(\ref{eqn:nonHermMarkovevo}) we get the QME of the system-plus-ancilla, 
\begin{equation}
\dot{\rhot} = \sL_{\rm sys}(\rhot) + \sD[ e^{-s/2} J_{1} \otimes 1 ](\rhot) + \sD[ \sqrt{2 S_{\rm l}} \otimes \sigma](\rhot) .
\label{maps}
\end{equation}
For $s>0$ we have $\alpha=0$, the mapping is efficient, and $\sqrt{2 S_{\rm l}} = \sqrt{1-e^{-s}} J_{1}$.  For $s<0$, in contrast, $\alpha \neq 0$ and the mapping is inefficient.
Alternative mappings to Eq.~(\ref{maps}) are obtained by exploiting the gauge invariance which may prove more convenient than Eq.~(\ref{maps}) for efficient numerical simulation. 

As an example of a system whose LD function can be observed with a system-plus-ancilla which corresponds to actual physical hardware, we consider the micromaser \cite{Englert2002}, an optical cavity pumped by excited two level atoms interacting with a thermal bath.  The micromaser has four distinct jump operators, $J_{1}=\sqrt{r} a^\dag \sin(\vartheta \sqrt{a a^\dag})/\sqrt{r a a^\dag}$ and $J_{2}= \sqrt{r} \cos(\vartheta \sqrt{a a^\dag}/\sqrt{r})$ corresponding to the observation of output atoms in the ground and excited states, respectively, and $J_{3}=\sqrt{\nu+1} a$ and $J_{4}=\sqrt{\nu} a^\dag$ associated to emission and absorption of quanta from the thermal bath.  
The micromaser has a rich dynamical phase diagram \cite{Garrahan2011,Horssen2012}, and in particular it displays multiple transitions in $\theta(s)$ as a function of $s$, when $s$ is the counting field conjugate to the number of jumps due to $J_{1}$, i.e.\ the number of outgoing atoms that have ceded a quantum to the cavity.
(Strictly speaking, since the micromaser is a few body system its dynamical transitions are actually sharp crossovers.)

We couple the micromaser to a two-level ancilla with weight operator $w = |1\rangle \langle 0|$, i.e.\ scheme (ii).  The QME which maps under $w$ to the corresponding $s$-ensemble, $\sW_{s}$ (at $s>0$), follows from Eq.~(\ref{eqn:notMarkovevo}),
\begin{equation}
\dot{\rhot} = \sL_{\rm sys}(\rhot) + \frac{1}{2} \sD[ J_{1} \otimes U_+ ](\rhot)
+ \frac{1}{2} \sD[ J_{1} \otimes U_- ](\rhot) ,
\label{micromaser}
\end{equation}
where $U_{\pm} \equiv e^{i\phi_\pm} p_{0} + p_{1}$, and $\phi_{\pm} \equiv \pm \cos^{-1}{(e^{-s})}$.
To obtain Eq.~(\ref{micromaser}) we have exploited the gauge invariance of Eq.~(\ref{eqn:notMarkovevo}). 
Note that the coupling to the ancilla is through the unitaries $U_{\pm}$, and so it can be achieved by using feedback, or simply scattering the outgoing quanta from the system off the ancilla. This is shown in Fig.\ \ref{fig1} with micro maser parameters $\vartheta=4\pi$, $\nu = 1$ and $r=1000$ \cite{Garrahan2011,Horssen2012}.

From Eq.~(\ref{eqn:quantweight}) we have that $\rhosys(t) = \Tr_{\rm a}[w_{t} \rhot(t)]$, so by measuring the time dependence of $w_{t}$, i.e. the coherence of the ancilla, we obtain the LD function, $\langle w_{t} \rangle = \Tr_{\rm s}[\rhosys(t)]\sim e^{t \theta(s)}$.  This means that from the rate of relaxation of $\langle w_{t} \rangle$ in the system-plus-ancilla we obtain the LD function of the system at a value of $s$ determined by the coupling to the ancilla through $U_{\pm}$.  Figure \ref{fig1} shows what would result from a quantum jump Monte Carlo simulation of the micromaser coupled to the ancilla.  From the rate of decay of $\langle w_{t} \rangle$ we obtain $\theta(s)$: the LD function estimated in this way displays a first-order singularity at $s_{c} \gtrsim 0$, as expected for the parameters of the figure \cite{Garrahan2011,Horssen2012}.

Such mappings to physical systems will always be possible when $s>0$. Thus a physical realisation for $s$ can be experimentally generated in many circumstances. Furthermore the large deviation function can be measured by doing measurements on the ancilla alone which need be no larger than a two level system. When $s<0$ efficient sampling is not possible. This restriction is a consequence of the norm growth, Eq.~(\ref{Trs}), when $s<0$. Efficient sampling of a probability distribution with growing norm embedded in a system with fixed norm is inherently impossible. Nonetheless, we expect there still exists a numerical advantage to the mapping, for any value of $s$. 

For classical many-body systems two numerical approaches have been used to sample rare trajectories. However, both these methods have seen limited application in open quantum many-body systems.  The first one is that of ``cloning'' (for a review see \cite{Giardina:2011}), where the non-conservation of probability [which is determined by the LD function $\theta(s)$) can be obtained by simulating in parallel a large number of clones of the system of interest.  Each clone evolves according to the unbiased (i.e., $s=0$) dynamics, with the addition of an interaction between clones, such that clones are either removed or duplicated with a probability that is dictated by the dynamical observable whose large-deviations one wishes to compute. In practice one has to reweigh the population to keep a constant number of clones, and from this reweighing factor the LD function is estimated \cite{Giardina:2011}.  Unfortunately, simulation of clones is vastly more numerically expensive in the quantum case as the total system size scales exponentially with the number of clones (as opposed to linearly in the classical).  

A second method is one based on transition path sampling \cite{Bolhuis:2002}. This is in essence a Monte Carlo scheme that performs a biased random walk in the space of trajectories, eventually converging to the a stationary distribution that corresponds to that of the leading eigenstate of the deformed master operator.  While this scheme is in principle generic, it is in practice efficient only when the dynamics obeys detailed balance, a condition that often does not apply in open quantum problems.  

Except for one recent exception \cite{Budini:2014}, neither TPS nor clones has been applied to estimate LD functions in quantum systems. In contrast the quantum weighting approach presented here has the advantage that it was designed using quantum theory from step one. Once a mapping has been found, the Lindblad master equation can be solved using numerical techniques previously developed for quantum systems: The spectral properties of quantum Lindblad operators are well known \cite{Kraus2008,Schirmer2010}, allowing direct diagonalization techniques to be optimised, or stochastic trajectories can be used to simulate the system, which provides both additional physical intuition and computational efficiency \cite{Gardiner2004,Wiseman2010}.

\section{ Application II: Inefficient Tracking Constraint} 

A quantum filter allows an experimentalist to make an optimal estimate of the quantum state given the record of a continuous homodyne measurement. For example, consider a quantum system evolving under a known Hamiltonian $H_p$, coupled to the environment with some operator $L_p$. The environment is measured using a homodyne detector at an angle $\phi$, which produces a stochastic continuous signal $y_t$. This signal contains information about the system observable $L_p e^{i\phi}+L_p^\dag e^{-i\phi}$ and can be used to estimate the current state through the equation
\begin{align}
d\pi_\phi = & (-i[H_p,\pi_\phi] + \sD[L_p](\pi_\phi)) dt \nn \\ 
&+ (L_p \pi_\phi e^{i\phi} +\pi_\phi L_p^\dag e^{-i\phi}) dy_\phi, \label{eqn:filterhomo}
\end{align}
where $\pi_\phi$ is an unnormalised density matrix which encodes the optimal estimate for the state and $dy_\phi$ is the change in the measurement signal over a time $dt$  \cite{Gardiner2004,Wiseman2010} (The above should be understood as a quantum Ito stochastic differential equation). Optimal estimates for observables can be calculated using $\langle X \rangle = \Tr[X \pi_{\phi/j}]/\Tr[\pi_{\phi/j}]$.  

Here we consider realising a quantum filter with Markovian quantum components in our framework. There are two reasons why this approach could have an advantage over simulating a quantum filter on a classical computer or circuitry. First, as the number of subsystems gets larger, integrating Eq.~(\ref{eqn:filterhomo}) on classical hardware rapidly becomes impractical due to the exponential growth of the Hilbert space dimension.  One might hope that this could be overcome by using instead quantum hardware \cite{Ortiz2001,Schindler2013,Bacon2001,Abrams1997,Aspuru-Guzik2012,Blatt2012,Bloch2012} to simulate such an equation.  Second, in principle a quantum filter made of quantum hardware could be integrated more easily into the microscopic scale of the quantum system it controls. This would also make the time scales of the filter and quantum system similar, ensuring the filter could provide an estimate sufficiently fast to correctly control the quantum system. However, we will show in our framework that creating a quantum filter using quantum hardware will always be inefficient. 
 
Note Eq.~(\ref{eqn:filterhomo}) is an alternative form to the one more commonly seen in the literature for quantum filters. These are often formulated in a normalised form in terms of some underlying stochastic process \cite{Gardiner2004,*Wiseman2010}. Specifically, Eq.~(\ref{eqn:filterhomo}) is mathematically equivalent to
\begin{align}
d\bar{\pi}_\phi = & (-i[H_p,\bar{\pi}_\phi] + \sD[L_p](\bar{\pi}_\phi)) dt + \bH[L_p e^{i\phi}](\bar{\pi}_\phi) dW, \label{eqn:normfilterhomo}
\end{align}
where $\bH[X](\cdot) = X \cdot + \cdot X^\dag - \langle X + X^\dag \rangle$,  $\bar{\pi}_\phi = \pi_\phi/\Tr[\pi_\phi]$. While the normalised and unnormalised forms are essentially the same, there is a fundamental difference in terms of which of the two is realisable with Markovian quantum hardware. Eq.~(\ref{eqn:normfilterhomo}) is nonlinear with regard to the quantum state.  Since it is \emph{impossible} to deterministically realise nonlinear evolution with Markovian quantum hardware \cite{Nielsen2010}, Eq.~(\ref{eqn:normfilterhomo}) is not realisable in this manner.  In contrast, the unnormalised equation, Eq.~(\ref{eqn:filterhomo}), acts linearly on the state, thus it is at least \emph{possible} in principle to realise it in terms of quantum hardware. 

The unnormalised filer Eq.~(\ref{eqn:filterhomo}) is a TLME [cf.\ Eq.~(\ref{eqn:genTLME}) with time dependent rates].  It is then possible, using the general framework described in the previous sections, to 
to construct an enlarged system with Markovian quantum components which will track the system of interest. We now show that such tracking will be inherently inefficient in the sampling sense of Sect.\ II.E.
To demonstrate this, we simply apply the efficiency test, Eq.~(\ref{eqn:repofalpha}), derived in Section \ref{sec:physrel} to Eq.~(\ref{eqn:filterhomo}). Technically, we can not do this directly, as we have assumed normal calculus applies when we derived the efficiency test. To circumvent this issue we change the stochastic integrals in Eq.~(\ref{eqn:filterhomo}) from Ito form to Stratonovich, giving us
\begin{align}
d\pi_\phi = & \big(-i[H_p,\pi_\phi] - \frac{1}{2} (L_p^\dag L_p + L_p^2 e^{2i\phi}) \pi_\phi \nn \\
& - \frac{1}{2} \pi_\phi (L^\dag_p L_p + (L_p^\dag)^2 e^{-2i\phi} )\big) dt \nn \\
& + ( L_p \pi_\phi e^{i\phi} + \pi_\phi L_p^\dag e^{-i\phi}) \circ dy_\phi.  \label{eqn:stratfilterhomo}
\end{align} 
Stratonovich integrals are compatible with regular calculus (but sacrifice the averaging property Ito integrals have). We take $B = (-i H_p - L^\dag_p L_p/2 + L^2_p e^{2i\phi}/2) dt + L_p e^{i\phi} \circ dy_\phi$,  $C = (i H_p - L^\dag_p L_p/2 + (L_p^\dag)^2 e^{-2i\phi}/2) dt + L_p^\dag e^{i\phi} \circ dy_\phi$, $D = 0$ and $E=0$. Replacing these matrices into Eq.~(\ref{eqn:repofalpha}) we find
\begin{align}
\alpha(t) = & \lambda_{max}^{+}\Big[-(L^2_p e^{2i\phi} +(L_p^\dag)^2 e^{-2i\phi} ) dt \nn \\
& + 2(L_p e^{i\phi} + L_p^\dag e^{-i\phi}) \circ dy_\phi \Big]. \label{eqn:alphatesthomo}
\end{align}
For efficient sampling to be possible we require $\alpha(t) =0 \; \forall \; t$. This is clearly not guaranteed to be the case for a general $L_p$. Thus a quantum filter realised with Markovian hardware using our framework will always have inefficient sampling.

There is a caveat: we can guarantee an efficient sampling if $L_p$ has the property $L_p^\dag = -L_p e^{i2\phi}$. However, this corresponds to a special case for the quantum filter where the system is no longer truly tracked. When $L_p^\dag = -L_p e^{i2\phi}$ the evolution of the filter becomes \emph{independent} of the evolution of the system that is being monitored. This is because the signal no longer contains information about the system, as the observable we are now measuring is $L_p e^{i\phi} + L_p^\dag e^{-i\phi} = 0$. Thus the filter becomes effectively decoupled from the evolution of the system. This means that if the filter and the system are started with precisely the same initial condition, the filter will follow the evolution of the system, but only because the measurement gives us the noise that the system is experiencing. However, if the filter is started with a different initial condition to the system, it will \emph{not} strictly converge to the system state \cite{Szigeti2014}. In this case the filter is not truly tracking the evolution of the system.

We have shown within our framework that quantum filters cannot be sampled efficiently, so the obvious question is: How can one escape the assumptions of our framework and find an experimental realisation with efficient sampling? To answer this question we first note: the coefficient $\alpha(t)$ directly relates to the growth of the norm of $\rhosys$, thus the inefficiency we observe is a direct consequence of the norm increasing. An analogous effect occurs in classical systems when one attempts to track the evolution of one system using another system with an equal number of degrees of freedom. The solution in the classical case is to instead use an \emph{ensemble} of weighted systems: so-called \emph{particle-filters} \cite{Arulampalam2002,*Rebeschini2013}. We suggest a similar approach may be appropriate in the quantum case: an ensemble of quantum systems with appropriately sampled quantum weights could be used to efficiently track a quantum system undergoing continuous measurement. Particle-filter-like techniques have already been applied to the simulation of quantum systems on classical computers \cite{Jacobs2010,Hush2013,*Hush2009}, and we suggest extension of these techniques could be used on quantum hardware. This particle filter approach escapes the restrictions of our framework because the quantum filter is composed of multiple quantum systems, in contrast to the above demonstration of inefficiency which assumed only one quantum system (with an attached ancilla) being used to track the target system.

\begin{acknowledgments} 
\noindent
This work was supported by EPSRC Grant no. EP/I017828/1. IL acknowledges funding from the European Research Council under the European Union's Seventh Framework Programme (FP/2007-2013) / ERC Grant Agreement No. 335266 (ESCQUMA).
\end{acknowledgments} 

\appendix

\section{System co-dependence} \label{apx:lessassumptions}

When deriving Eq.~(\ref{eqn:preweightalgebra}) we used an ansatz: each term in Eq.~(\ref{eqn:wgenMarkov}) (repeated below) individually satisfied the closure property.
\begin{align}
\dot{\rhosys} =& \sLsys(\rhosys) + \alpha(t) \rhosys +  e^{\int_0^t ds \; a(s)} \Bigg( \sum_j^{J_c} ( -iM_j \Tr_a[w f_j \rhotot] \nn \\
&-iM^\dag_j \Tr_a[w f^\dag_j \rhotot] +i\Tr_a[f_j w \rhotot] M_j +i \Tr_a[f^\dag_j w \rhotot] M^\dag_j) \nn \\
& + \frac{1}{2}  \sum_{i,j}^{J_d} \sum_{k}^{K_d} ( 2N_{i,k} \Tr_a[ g_{j,k}^\dag w g_{i,k} \rhotot ] N_{j,k}^\dag \nn \\
& -N_{j,k}^\dag  N_{i,k}\Tr_a[ w g_{j,k}^\dag g_{i,k}   \rhotot ] \nn \\ 
&-\Tr_a[ g_{j,k}^\dag g_{i,k} w \rhotot ] N_{j,k}^\dag N_{i,k}) \Bigg).  \label{eqn:repwgenMarkov}
\end{align}
Here we consider weakening this assumption, and briefly justify why the theorem still holds even when multiple terms are grouped together and are required to satisfy the closure property. 

Consider the case where we set $M_j^\dag = M_j$ in this case Eq.~(\ref{eqn:repwgenMarkov}) becomes
\begin{align}
\dot{\rhosys} & = e^{\int_0^t ds \; a(s)} \Big( \sum_j^{J_c} ( -i M_j \Tr_a[w (f_j+f^\dag_j) \rhotot] \nn \\
& + i\Tr_a[(f_j +f^\dag_j) w \rhotot] M_j + \cdots .  \label{eqn:changedwgenMarkov}
\end{align}
The first and second terms produce the requirement $w (f_j+f^\dag_j) = \gamma_j^{\rm r}$ and $(f_j+f^\dag_j) w= \gamma_j^{\rm r}$. Applying the Moore-Penrose inverse in a method identical to what was presented in section \ref{eqn:formofsupG} it is straightforward to show that $\gamma_j^{\rm l/r}=(\gamma_j^{\rm l/r})^*$ meaning both these constants are real. Replacing these identities into Eq.~(\ref{eqn:changedwgenMarkov}) we find
\begin{align}
\dot{\rhosys} & = e^{\int_0^t ds \; a(s)} \Big( \sum_j^{J_c} ( -i \gamma_j^{\rm l} M_j \rhosys + i \gamma_j^{\rm r} \rhosys M_j + \cdots .
\end{align}
Both $ -i \gamma_j^{\rm l} M_j $ and $i \gamma_j^{\rm r} \rhosys M_j$ are \emph{anti-hermitian} operators. Thus the form for the superoperator $\sM(\rhosys)$, as given in Eq.~(\ref{eqn:effnotMarkovevo}), is not modified by placing restrictions on the system operator $M_j$. 

Another option is to set $ M_{j'} = i N_{j,k}^\dag N_{i,k}$, where we set $J_c = J_d^2K_d$ and $j' = i+J_d(j-1) + J_d^2(k-1)$. Replacing into Eq.~(\ref{eqn:repwgenMarkov}) gives
\begin{align}
\dot{\rhosys} & = e^{\int_0^t ds \; a(s)} \Big( \sum_{i,j}^{J_d} \sum_{k}^{K_d} -N_{j,k}^\dag N_{i,k} \Tr_a[w (-f_{j'} + g_{j,k}^\dag g_{i,k}) \rhotot]  \nn \\
& - N_{i,k}^\dag  N_{j,k} \Tr_a[w f_{j'}^\dag \rhotot] - \Tr_a[ (f_{j'} + g_{j,k}^\dag g_{i,k}) w \rhotot] N_{j,k}^\dag  N_{i,k} \nn \\
& + \Tr_a[w f_{j'}^\dag \rhotot] N_{i,k}^\dag  N_{j,k} +\cdots .
\end{align} 
These terms produce the following modified quantum weight algebra: when $i\ne j$ $w (-f_{j'} + g_{j,k}^\dag g_{i,k}) = {\kappa^{\rm r}_{i,j,k}}'w$, $ (f_{j'} + g_{j,k}^\dag g_{i,k}) w = {\kappa^{\rm l}_{i,j,k}}'w$, $wf_{j'}^\dag = (\gamma^{\rm r}_{j'})^*w$ and $f_{j'}^\dag w = (\gamma^{\rm l}_{j'})^*w$; and when $i=j$: $w (-f_{j'} + f_{j'}^\dag + g_{j,k}^\dag g_{i,k}) = {\kappa^{\rm r}_{i,j,k}}'w$ and $ (f_{j'} -f_{j'}^\dag + g_{j,k}^\dag g_{i,k}) w = {\kappa^{\rm l}_{i,j,k}}'w$. 

Applying methodology from section \ref{eqn:formofsupG}, and taking special care to split equations into their Hermitian and anti-Hermitian parts where possible it can be shown: $w g_{j,k}^\dag g_{i,k} = \kappa^{\rm r}_{i,j,k}w$, $g_{j,k}^\dag g_{i,k} w = \kappa^{\rm l}_{i,j,k}w$, $w f_{j'} = \gamma^{\rm l}_{j'}$, $w f_{j'}^\dag = (\gamma^{\rm l}_{j'})^*$, $f_{j'} w = \gamma^{\rm l}_{j'}$, $f_{j'}^\dag w = (\gamma^{\rm l}_{j'})^*$. Where ${\kappa^{\rm r}_{i,j,k}}' =\kappa^{\rm r}_{i,j,k} - \gamma^{\rm r}_{j'}$, ${\kappa^{\rm l}_{i,j,k}}' =\kappa^{\rm l}_{i,j,k} + \gamma^{\rm r}_{j'}$ and $\Re[\gamma^{r/l}_{j'}]=0$ when $i=j$. We have successfully split the modified quantum algebra, into a form which is the same as Eq.~(\ref{eqn:preweightalgebra}), thus the proof continues in an almost identical manner to what is presented in section \ref{eqn:formofsupG}. The only caveat being the  is when $i=j$ in which case the purely imaginary property of $ \gamma^{r/l}_{j'}$ produces anti-Hermitian terms in an identical manner to the case previous case (Eq.~(\ref{eqn:changedwgenMarkov})). Thus the form for the superoperator $\sM(\rhosys)$, as given in Eq.~(\ref{eqn:effnotMarkovevo}), is not modified by setting $ M_{j'} = i N_{j,k}^\dag N_{i,k}$. 

\section{Operator independence} \label{apx:complicatedNepdellr}

In this appendix we discuss how an ancilla-coupling can be chosen such that the operators:
\be
N_k^{\delta/\epsilon,\rm l/r} = \sum_i^{J_d} N_{i,k} | (\delta/\epsilon)^{\rm l/r}_{i,k} ), \label{eqn:defnNdeleplr}
\ee
are all independent and operate on the system space. The first choice we make is that the vectors $\delta^{\rm l/r}_{i,k}$ and $\epsilon^{\rm l/r}_{i,k}$ are both of dimension 1. Furthermore we assume that the singular values of $w$ are all the same, in this case the constants $\delta^{\rm l/r}_{i,k}$ and $\epsilon^{\rm l/r}_{i,k}$ become independent (as explained after Eq.~(\ref{eqn:proofofkappal})). Ancilla-system coupling exist where this is the case, indeed Eq.~(9) and (10) from the main text are both examples of this.

Next we split the index over $i$ into four parts and assign the following values for $\delta^{\rm l/r}_{i,k}$ and $\epsilon^{\rm l/r}_{i,k}:$ when $i \in \mJ_d^{\delta, \rm l} = [1, J_{\delta, \rm l}]$ we set $\delta^{\rm l}_{i,k} =1$ and $(\delta^{\rm r}/\epsilon^{\rm l/r})_{i,k} =0$; when  $i \in \mJ_d^{\delta, \rm r} = [J_{\delta, \rm l} +1, J_{\delta, \rm r}]$ we set $\delta^{\rm r}_{i,k} =1$, $(\delta^{\rm l}/\epsilon^{\rm l/r})_{i,k}=0$; when $i \in \mJ_d^{\epsilon, \rm l} = [J_{\delta, \rm r}+1, J_{\epsilon, \rm l}]$ we set $\epsilon^{\rm l}_{i,k} =1$, $(\delta^{\rm l/r}/\epsilon^{\rm r})_{i,k}=0$; and $i \in \mJ_d^{\epsilon,\rm r} = [J_{\epsilon,\rm l}+1, J_{\epsilon,\rm r}]$ we set $\epsilon^{\rm r}_{i,k} =1$, $(\epsilon^{\rm l}_{i,k}/\delta^{\rm l/r})_{i,k} =0$. Replacing these definitions into Eq.~(\ref{eqn:defnNdeleplr}) we find get 
\be
N_k^{\delta/\epsilon,\rm l/r} = \sum_{i \in \mJ_d^{\delta/\epsilon,\rm l/r}} N_{i,k} , \label{eqn:indpenNdeleplr}
\ee
where we dropped the inner product notation, as the inner product reduces to the regular product when the dimension of the vector is 1. The sets $\mJ_d^{\delta/\epsilon,\rm l/r}$ are all disjoint by construction. Thus $N_k^{\delta/\epsilon,\rm l/r}$ are all independent operators.

\bibliography{HiddenQuantumMarkovReferences}

\end{document}